\documentclass{pnastwo}

\usepackage{amssymb,amsfonts,amsmath}
\usepackage{epsfig}
\usepackage{color}

\newcommand{\4}{$_{4}$}
\newcommand{\cm}{cm$^{-1}$}
\newcommand{\ai}{\textit{ab initio}}
\newcommand{\Td}{${\mathcal T}_{\rm d}$}
\newcommand{\um}{$\mu$m}

\contributor{Accepted by Proceedings
of the National Academy of Sciences of the United States of America}
\url{www.pnas.org/cgi/doi/10.1073/pnas.1324219111}
\copyrightyear{2014}

\begin{document}



\title{The spectrum of hot methane in astronomical objects
using a comprehensive computed line list}





\author{Sergei N. Yurchenko,\affil{1}{
Department of Physics and Astronomy, University College London, London WC1E 6BT, UK}
Jonathan Tennyson,\affil{1}{}
Jeremy Bailey,\affil{2}{School of Physics, University of New South Wales, NSW 2052, Australia}
Morgan D. J. Hollis,\affil{1}{}
\and
Giovanna Tinetti\affil{1}{}
}

\contributor{Accepted by Proceedings of the National Academy of Sciences of the United
States of America}

\footlineauthor{Yurchenko, Tennyson, Bailey, Hollis, Tinetti}

\maketitle

\begin{article}

\begin{abstract}

Hot methane spectra are important in environments ranging from
flames to the atmospheres of cool stars and exoplanets. A new spectroscopic line list,
10to10, for
$^{12}$CH$_4$
  containing almost 10 billion transitions is presented. This
  comprehensive line list covers a broad spectroscopic
  range and is applicable for temperatures up to 1\,500~K. Previous
  methane data are incomplete leading to
  underestimated opacities at short wavelengths and elevated
temperatures. Use of 10to10 in models of the
bright T4.5 brown dwarf 2MASS 0559-14
 leads to
significantly better agreement with observations and in studies
of the hot Jupiter exoplanet HD~189733b leads to up to a twentifold increase in methane
abundance. It is
demonstrated that proper inclusion of the huge increase
in hot transitions which are important at elevated
temperatures is crucial for accurate characterizations of
atmospheres of brown dwarfs and exoplanets, especially when
observed  in the near-infrared.

\end{abstract}







\section{Significance Statement}

Hot methane is important in cool stars, brown dwarfs, exoplanets, gas turbine engines and elsewhere. There is a pressing need for an accurate and complete spectroscopic database for methane at elevated temperatures. We present a comprehensive spectroscopic line list for methane containing almost 10 billion transitions, 2\,000 times more than any previous compilation, covering a broad spectroscopic range and applicable for temperatures up to 1\,500 K. We demonstrate that such a line list is essential for correctly modelling the brown dwarf 2MASS 0559-14 and leads to large changes in exoplanet models. We believe that this line list will make a large impact on the field of exoplanets and cool stars.

\mbox{}\\

\dropcap{M}ethane is an important terrestrial green house gas
\cite{04RoDoxx.CH4} and the active component in many flames. It is
also the major active atmospheric constituent of cool carbon stars and
the T-dwarf class of brown dwarfs are often referred to as ``methane
dwarfs'' \cite{05Kixxxx.dwarfs,06BuGeLe.dwarfs}. Methane is also important for the newly discovered Y dwarfs \cite{11CuKiDa.dwarfs}, and even L dwarfs \cite{97NoGeMa.dwarfs}.
 Methane has been
detected in exoplanets \cite{08SwVaTi.exo,10SwDeGr.exo,11JaCaTh.exo},
where it has long been thought of as a potential biosignature
\cite{93SaThCa.CH4,07AtMaWo.CH4} on earth-like planets.  Equilibrium
chemistry models for exoplanet gas giants \cite{02LoFexx.exo} predict
methane to be the main carbon containing species at temperatures
below 1\,500 K, so measuring its abundance is essential for determining
the key C/O ratio in these objects \cite{07TiLiVi.exo,11MaHaSt.exo}.
Even though methane is therefore likely to be important in many of the giant
exoplanets detected so far, its observed abundance remains
controversial \cite{10StHaNy.exo,jt495}.
Studies of many topics, ranging from
the astronomical ones just mentioned to halon flame inhibitors
\cite{97McDaMi.CH4}, combustion \cite{07JoGaCh.CH4}, gas turbine
energies \cite{09DiRaJe.CH4} and exhausts \cite{01MaCoNe.CH4}, all
rely on an understanding of the spectroscopy of hot methane.  However
even at room temperature the spectrum of methane is complex and not
fully characterized \cite{13BrSuBe.CH4}. At elevated temperatures, above about 1000~K,
billions of  spectral lines become active and modelling hot
methane with available laboratory data has often
proved difficult \cite{96GeKuWi.CH4,12BaKexx.dwarfs}.  The importance
of an accurate and complete line list for methane has been stressed
many times \cite{07ShBuxx.dwarfs,08FrMaLo.exo,11BaAhMe.exo,jt528}.


None of the available CH\4\ line lists are complete at elevated
temperatures \cite{07ShBuxx.dwarfs}.  The recently updated CH\4\ data
\cite{13BrSuBe.CH4} included in the 2012 edition of HITRAN \cite{jt557}
is designed to work at Earth atmosphere temperatures.  Current high
temperature line lists include ones computed from first principles
\cite{09WaScSh.CH4} and semi-empirically \cite{13BaWeSu.CH4} as well
as (partial) experimental line lists
\cite{03NaBexx.CH4,08ThGeCa.CH4,12HaBeMi.CH4,13CaLeMo.CH4}.  As
demonstrated below, all of these compilations underestimate the number
of methane transitions that need to be considered at elevated
temperatures by many orders of magnitude.




\begin{figure}[h]
\vspace{-0.3cm}
\centering
{\leavevmode \epsfxsize=8.0cm \epsfbox{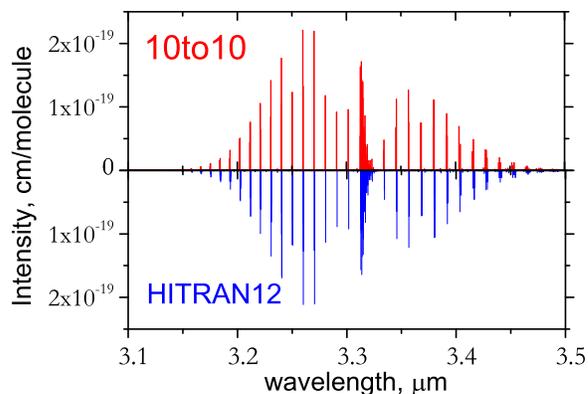}}
\caption{Absorption intensities of the $\nu_3$ band of $^{12}$CH\4\ at $T=296$~K: HITRAN (lower display)
 and 10to10 (upper display). \label{f:nu3}}
\end{figure}

Here we present a new methane line list computed variationally which
contains 9.8 billion transitions and is designed to be complete for
temperatures up to 1\,500~K. It covers  wavelengths from the
far-infrared to 0.9~$\mu$m. This can be compared to about 340\,000
transitions of $^{12}$CH$_4$ known experimentally for the
same wavelength region. We show that
proper inclusion of the huge increase in hot transitions which are
important at elevated temperatures is crucial for accurate
characterization of atmospheres of brown dwarfs and exoplanets,
especially when observed in the near-infrared.

\section{The 10to10 line list}

Methane  has 5 atoms and hence 9 vibrational modes, more than are
routinely treated using variational nuclear motion programs. However
it is a highly symmetric molecule belonging to the \Td\ point
group which allows some simplification.  The ro-vibrational energies,
associated wave functions and Einstein A coefficients, which are required
to generate spectra, were computed using a specifically adapted version of the
nuclear motion program TROVE \cite{07YuThJe.method} in conjunction with a new
spectroscopically obtained potential energy surface (PES) and a
previously-calculated, \ai\ dipole moment surface DMS \cite{jt555}.
The energies were computed variationally, i.e. by diagonalizing a
large set of (200) Hamiltonian matrices constructed using a suitable
basis set in the symmetry adapted representation.  The largest matrix
considered (for rotational state $J=39$) had $163\,034 \times
163\,034$ elements and was diagonalized using the (non-sparse)
ScaLAPACK MPI direct eigensolver PDSYEVR. A direct diagonalization of
large non-sparse matrices is the bottleneck in variational computation
of comprehensive line lists for polyatomic molecules. These
calculations were performed at the UK National Cosmology Supercomputer
COSMOS and required about 1.5M CPU hours. Calculating the line intensities is also computationally
expensive but more straightforward;
these calculations are independent and thus can be very efficiently
parallelized. This part of the project was performed at The Cambridge
High Performance Computing Cluster Darwin ($~$3.0M CPU hours).
Full details of the
computational method used to construct the line list will be published elsewhere \cite{14YuTexx.CH4}.


To reduce computational requirements, we imposed an upper energy
limit of $E = hc$ 18\,000~\cm. With this threshold for all rotational
states up to $J=39$, we obtained 6\,603\,166 energy levels.  Einstein A
coefficients were computed for all transitions involving rotational excitations
$J=0\ldots 39$ with the lower and upper state energies  ranging up to
$ hc$ 8\,000~\cm  and $hc$ 18\,000~\cm, respectively, and covering the wavenumber range up to 12\,000~\cm\ (0.83~\um).
A total of 9\,819\,605\,160  transitions were computed. The energy values,
transition wavenumbers, Einstein A coefficients and degeneracy factors
comprise our 10to10 line list which can be obtained from
\texttt{www.exomol.com}.

The large basis set used in the diagonalization plus the combination of an
accurate variational model, an empirical PES and \ai\ DMSs obtained at the high
level of \ai\ theory \cite{13NiReTy.CH4} all contribute to the high quality of the
10to10 line list. Further technical details are given in the
supporting material. This line list forms a key part of the ExoMol
project \cite{jt528}, which aims to provide such lists for all
molecules likely to be detected in exoplanetary atmospheres.

\section{Line list results}

Typical agreement with the experimental spectra is shown in
Fig.~\ref{f:nu3}, where the absorption intensities for the $\nu_3$
(stretching asymmetric) band at $T=296$~K are shown as  a stick diagram.
Not only do the theoretical line positions agree well with the
measurements, the absolute intensities reproduce the experimental data
with an accuracy comparable to  experiment, where the latter is available. Validation by
experiment is important because it gives us confidence in our
predictions of the methane opacity. Indeed, our intensities are based
purely on the quality of the \ai\ DMS, without any empirical
adjustments.

The new 10to10 line list offers at least 200 times more lines than any
other previous compilation, including the most complete: the
empirically constructed MeCaSDa line list \cite{13BaWeSu.CH4}. Most of
these extra transitions come from spectral lines which involve high
rotational levels and/or transitions between vibrationally excited
states (``hot bands''). These transitions are generally very weak in
room-temperature spectral but become important when the temperature
increases.
To demonstrate the importance of these transitions we have counted
the number of active transitions as a function of temperature.
At
room temperature only 1~\%\ of the 10to10 lines have intensities
stronger than $10^{-30}$~cm/molecule, which roughly corresponds to the scope of CH\4\ in the HITRAN~2012 database \cite{jt557}.
This shows that most 10to10 lines are unimportant for
room temperature applications. However at $T=1\,500$~K, 98~\%\ of
10to10 lines have absorption intensities stronger than
$10^{-32}$~cm/molecule, 80~\%\ are stronger than $10^{-30}$ and 40~\%\
are stronger than $10^{-29}$ cm/molecule.  Besides the density of
these lines is extremely high as 66\,000 lines per cm$^{-1}$ at $T=$
1\,500~K for the standard HITRAN cutoff of $10^{-29}$~cm/molecule. This
means that billions of transitions are required to model the high
temperature spectrum of methane.

\begin{figure}[h!]
\vspace{-0.3cm}
\centering
{\leavevmode \epsfxsize=7.0cm \epsfbox{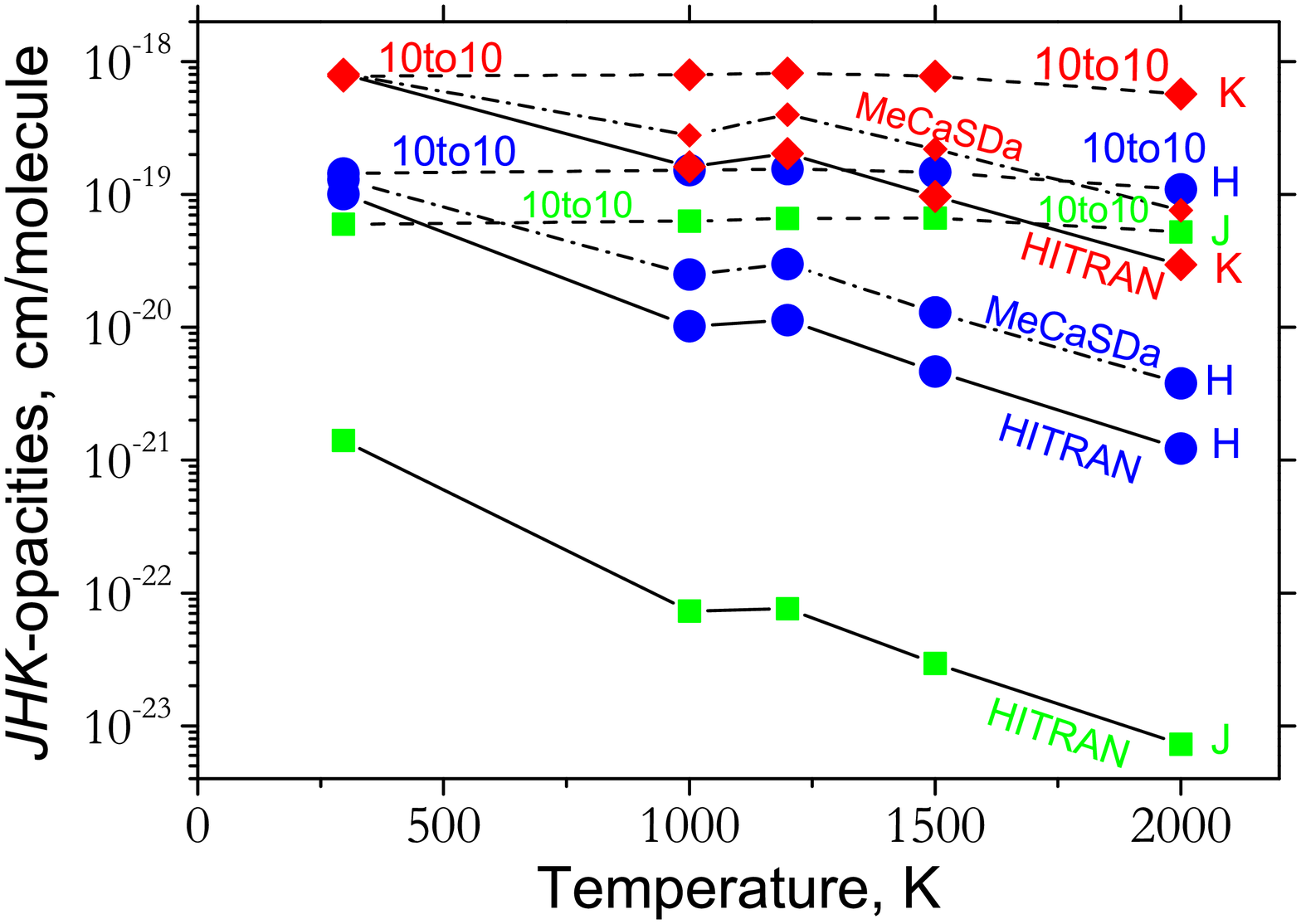}}
\caption{Integrated intensities for the $K$, $H$, and $J$ bands
computed using different line lists: HITRAN~2012 \cite{jt557}, MeCaSDa \cite{13BaWeSu.CH4} and 10to10. MeCaSDa gives the largest $K$ and $H$ coefficients for all the previously available line lists but does not cover the $J$ band, for which all previous
line lists are very incomplete even at room temperature.
\label{f:JHK}}
\end{figure}

It is common astronomical practice to estimate the integrated flux
within so-called $J,H,K,L,M \ldots $  colors, or bands, which correspond to the transparency windows of the Earth atmosphere in
the infrared. Here we present integrated opacities of CH\4\
at different temperatures
for the three main methane spectroscopic windows: the
$J$, $H$, and $K$ bands which are defined as
1.1--1.4~$\mu$m, 1.5--1.8~$\mu$m, and 2.0--2.4~$\mu$m, respectively.
Fig.~\ref{f:JHK} compares 10to10 integrated intensities with estimates
obtained from the HITRAN and MeCaSDa databases.  Agreement is good at room
temperature for the $H$ and $K$ but {\it all} previous compilations
show a rapid drop in the methane opacity with temperature, while
10to10 suggests it is approximately flat.
10to10 gives enhanced
absorption in the $J$ band for all temperatures which is to be
expected as this region is poorly sampled experimentally
\cite{13BrSuBe.CH4}.
At high temperatures ($T>1\,500$~K), however, the 10to10 intensities deviate from a flat line,
showing incompleteness of the 10to10 line list for such temperatures. In order to improve the temperature
coverage to, say, 2\,000~K, the lower energy threshold has to be also extended at least up to 10,000~\cm\ and
the rotational excitations to about $J=45$. This work is currently underway.

\begin{figure}[h!]
\centering
\vspace{-0.5cm}
{\leavevmode \epsfxsize=7.0cm \epsfbox{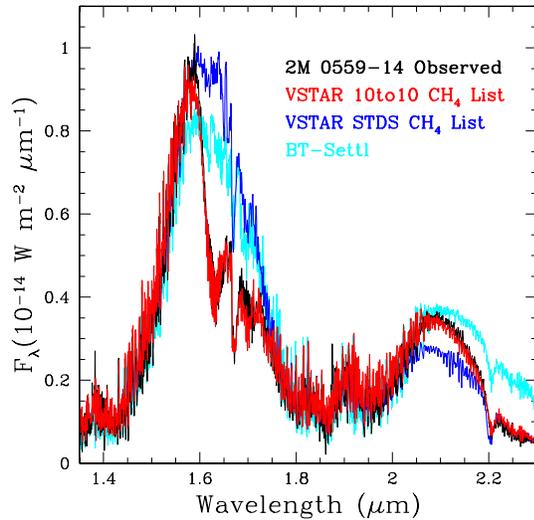}}
\caption{VSTAR spectra of the T~4.5 dwarf: models are VSTAR with the 10to10 line list (this work),
VSTAR with the STDS line list \cite{12BaKexx.dwarfs} and a comparison using the BT-Settl model
of of Allard {\it et al} \cite{07AlAlHo.exo}.
The observed spectrum was taken with the SpeX instrument on the 3 m NASA Infrared Telescope Facility (IRTF) and
was obtained from the IRTF spectral library \cite{05CuRaVa.dwarfs,09RaCuVa.dwarfs}. It has a spectral resolving power
($R$ = $\lambda/ \Delta \lambda$) of 2000, and a S/N of better than 50.
\label{f:T4.5}}
\end{figure}

\section{Astronomical models}


In Fig.~\ref{f:T4.5} we present the results of modelling a spectrum of
the bright T4.5 dwarf 2MASS 0559-14 taken from the IRTF spectral
library \cite{05CuRaVa.dwarfs,09RaCuVa.dwarfs}.
At this spectral type methane bands at around 1.6 and 2.2 \um\ are becoming prominent features in the spectrum.
Previous analyses of the spectrum of this object have derived effective temperatures in the range 1\,000--1\,200~K and
gravities of $\log g$ (in cgs units) = 4.5 - 5.0  \cite{09StLeCu.dwarfs,09deMaZa.dwarfs}.
 The model spectrum was
calculated using the VSTAR code and the methods described by Bailey
\&\ Kedziora-Chudczer \cite{12BaKexx.dwarfs}. The model is based on a
pressure temperature structure for effective temperature 1\,200 K and
$\log g = 5$ \cite{06BuSuHu.dwarfs} and assumes equilibrium chemistry.
The new
model also uses updated absorption coefficients for the H$_2$--H$_2$
collision induced absorption (CIA) \cite{11AbFrLi.H2}.
The blue spectrum on Figure~\ref{f:T4.5} corresponds to the previous
model which used the STDS methane line list \cite{98WeChxx.CH4}, a
precursor to the recent MeCaSDa line list \cite{13BaWeSu.CH4}.
The red spectrum is the new model using the 10to10 line list: The comparison of the new VSTAR model using 10to10
with the old VSTAR model using the STDS list provides the direct comparison of the effect of just changing the opacities.
Also included on the plot are results obtained using the BT-Settl model
\cite{07AlAlHo.exo} for $T_{\rm eff} = 1\,200$ K and $\log g = 5$ which is also based
on the old line lists and similarly fails to match the observations in the regions of strong methane absorption

\begin{figure}[h!]
\vspace{-0.3cm}
\centering
{\leavevmode \epsfxsize=7.5cm \epsfbox{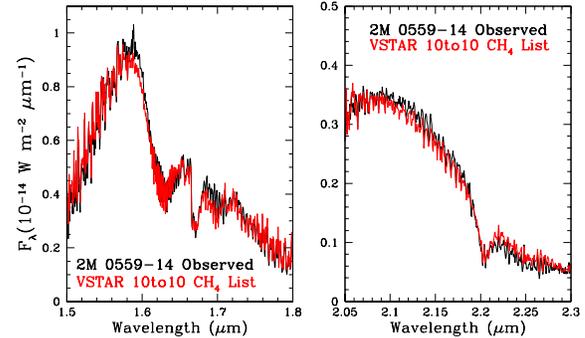} }
\caption{Close up of spectra of the T~4.5 dwarf observed using IRTF compared
to a  VSTAR model with the 10to10 line list. \label{f:T4.5close}}
\end{figure}

It can be seen that the new model using the 10to10 list fits the data
much better, particularly in the 1.6 to 1.8 $\mu$m region. In this
region the STDS based lists used previously include no hot bands, and
clearly fail to properly represent the absorption in this region.
Figure~\ref{f:T4.5close} shows two expanded views of the spectral
regions sensitive to methane and shows that much of the detailed line
structure is reproduced by the model. The VSTAR model used here, which
is a cloud free model, fits the data for this object significantly
better than the BT-Settl model which include clouds.

A line list as large as 10to10 presents something of a challenge for a
line-by-line modelling code such as VSTAR.  Initially we tried to
model the spectrum using a line intensity cutoff of 10$^{-27}$ cm/molecule
at 1500 K to reduce the number of lines used in the line-by-line code.
However, this procedure was found to produce significantly less
absorption than the use of the full line list. Even though the lines
below this cutoff are very weak, the very large number of them means
they still contribute significant total absorption.  In order to
speed up the line-by-line calculations we have divided the lines into
strong and weak lines. Only the strong methane lines are modelled with
a full line shape calculation. For the many weak lines we simply add
the total line absorption into the single wavelength bin in which the
line center lies. Given that there are many more lines than there are
wavelength bins in our spectrum these lines are clearly never going to
be individually resolved. In the models presented here the boundary
between strong and weak lines was set at $10^{-26}$ cm/molecule at the
temperature of the atmospheric layer being modelled.




\begin{figure}[h!]
\vspace{-0.5cm}
\centering
{\leavevmode \epsfxsize=9.5cm \epsfbox{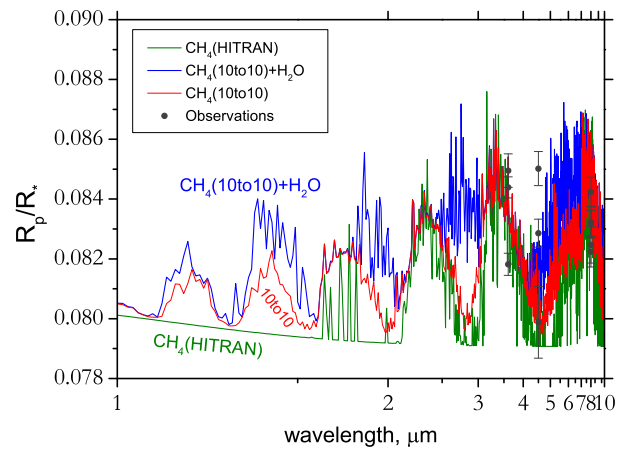}}
\caption{\textsc{tau} transmission spectra of the atmosphere of exoplanet GJ~436b
at 700~K: 10to10 spectrum (methane only), HITRAN spectrum (methane only), and CH\4(10to10)+water. Observations data are from Beaulieu et al. \cite{jt495}
and Knutson et al. \cite{11KnMaCo.exo}. A mixing ratio of
$10^{-5}$ both for H\2O and CH\4 was used. Note that new observations in the 1.2 -- 1.7 $\mu$m region indicate the presence of  clouds or hazes \cite{14KnBeDe.exo} which are not included in our model.
\label{f:GJ436b} }
\end{figure}

It remains very challenging to obtain even relatively crude spectra
of exoplanets. The most productive technique has been monitoring the
variation in observed starlight for those planets that transit their
host star as viewed from Earth, for both the primary and secondary
eclipse. Interpreting these observations requires the construction
of detailed radiative transfer models.
\textsc{tau}  \cite{jt521,13HoTeTi.exo} is a one-dimensional radiative
transfer code for transmission spectroscopy of exoplanets, especially
designed for hot Jupiters with near stellar orbits.
It uses a line-by-line integration scheme to
model transmission of the radiation from the parent star through the
atmosphere of the orbiting planet, equating physically to observations
of the radius ratio (transit depth) as a function of wavelength in the
primary transit geometry. Abundances of absorbing molecules in the
atmosphere can hence be estimated by hypothesizing compositions and
comparing to any available observations. Specifically, the algorithm
calculates the optical depth of the atmosphere (and hence effective
radius of the planet) at a particular wavelength, given model trace
molecular abundances and the atmospheric structure and absorbing
behavior (in the form of line lists) of those molecules. A transit
depth can hence be calculated as the ratio of the squared radii of the
planet and the star, and a spectrum created of absorption as a
function of wavelength.


\begin{figure}[h]
\vspace{-0.5cm}
\caption{\textsc{tau} transmission spectra of the atmosphere of exoplanet HD~189733b
at 1\,000~K: 10to10 spectrum (methane only),  experimental (HITRAN) spectrum (methane only), and the corresponding  mixed H\2O+CH\4\ spectra.
The observations  are a compilation of available measurements from Refs.~\cite{07TiViLi.exo,
07KnChAl.exo,
07TiLiVi.exo,
08SwVaTi.exo,
09SiDeDe.exo,09DeDeHe.exo,09KnChCo.exo,
10AgCoKn.exo,
11SiPoAi.exo,11DeSiVi.exo,
12GiAiPo.exo,12GiAiRo.exo,
13PoSiGi.exo,14DaDeTi.exo}.
A mixing ratios of  $5\times 10^{-4}$ both for H\2O and CH\4; the radius offset is 4.5\%.
\label{f:HD189} }
\centering
{\leavevmode \epsfxsize=9.5cm \epsfbox{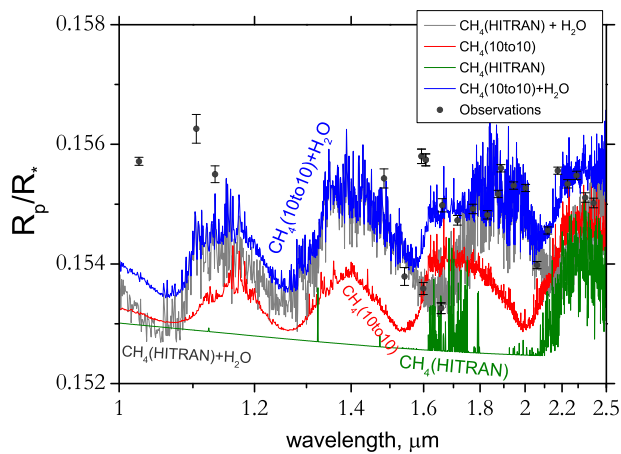}}
\end{figure}

We used \textsc{tau} \cite{jt521,13HoTeTi.exo} to generate models to
simulate the atmospheric transmission of the warm Neptune GJ~436b
\cite{GJ436b} and of the hot Jupiter HD~189733b \cite{HD189733b}, see
Figs.~\ref{f:GJ436b} and \ref{f:HD189} respectively. Both these
exoplanets have been extensively observed \cite{jt495, 11KnMaCo.exo,
  07TiViLi.exo, 07KnChAl.exo, 07TiLiVi.exo, 08SwVaTi.exo,
  09SiDeDe.exo,09DeDeHe.exo,09KnChCo.exo, 10AgCoKn.exo,
  11SiPoAi.exo,11DeSiVi.exo, 12GiAiPo.exo,12GiAiRo.exo,
  13PoSiGi.exo,14DaDeTi.exo,14KnBeDe.exo}.  Assuming for GJ~436b an equilibrium
temperature of about 700~K, the atmosphere of this planet is expected
to contain significant quantities of methane as a trace absorber in
chemical equilibrium \cite{99BuShxx.exo,13MoLiVi.exo,14VeAgSe.exo} as
well as some water. In models for GJ~436b containing only methane, the
differences in the line lists result in an observed theoretical radius
ratio increase $\Delta (R_p/R_*)$ of up to 0.3\%\ at 5~\um\ and below
2~\um\ (after smoothing).  Similar models on the well-studied
exoplanet HD~189733b, a hot Jupiter with an effective temperature
1\,200~K, found an increase of up to 0.6\%, see Fig.~\ref{f:HD189}. If
due solely to methane opacity, this would require an increase in
modelled abundance of approximately 5 times using the HITRAN opacities
to match the level predicted by 10to10. A large portion of these
regions is however screened by water.  This can be also seen in
Figs.~\ref{f:GJ436b} and \ref{f:HD189}, where the results of the
CH\4+H\2O models are presented.  In particular, 5~\um\ regions is
completely dominated by the water absorption making the effect from
the new methane data at this wavelength negligible.  However the
windows around 1.6~\um\ and 2.2~\um\ show a significant difference
with an increase of the radius ratio of up to 0.1~\%\ for the
CH\4(10to10)+H\2O model of the hot Jupiter HD~189733b in
Fig.~\ref{f:HD189}.  The corresponding mixing ratios of CH$_4$ and
H$_2$O were obtained by fitting to the observations available in the
literature. In addition to the 10to10 and HITRAN~2012 line lists for
methane, the BT2 line list \cite{jt378} was used to model
absorption by water.  It should be noted that the large uncertainties
of the existing observation data points, some of which appear to
contradict each other (see, e.g., Fig.~\ref{f:GJ436b}), lead to a very
large number of possible solutions.  Figures~\ref{f:GJ436b} and
\ref{f:HD189} show only one selection of models as an example
illustrating the significance of the potential errors involved in
using line lists that are known to have missing opacities.
These errors may not have been obvious in previous
studies, since the methane discrepancies are often masked to a large
extent by the inclusion of realistic quantities of water at around the
$10^{-5}$ abundance level.



\section{Conclusion}


We present a new comprehensive line list for methane suitable for
modelling absorptions up to 1\,500~K.  An important consideration is
that use of 10to10 increases opacities generated by methane which
changes the level of the atmosphere at which methane is absorbed or
emitted. Future exoplanet models will have to account for the new opacities to
retain self-consistency.  The increased methane opacities are expected
to have some effect on the pressure temperature profile and thermal
evolution of brown dwarfs and exoplanets. However this is beyond the
scope of the models considered here.  Such effects can be properly
evaluated when the new opacities are included in self-consistent
structure models for the atmospheres and interiors. However, we would
expect the changes to be relatively small as the extra opacity seen
with the new line list is a relatively small fraction of the total
opacity due to all sources, such as water, methane, CIA.

Despite the size of
the 10to10 line list it is still not complete. We can estimate
the effect of the incompleteness of the 10to10 line list by
comparison with
the high temperature partition function \cite{08WeChBo.CH4}. The
critical parameter here is the constraint introduced by our use of
a lower energy threshold at $hc
\, 8\,000$~\cm\ on the states that can be thermally occupied.
The partition function of CH\4\ computed using 10to10 levels lying
below this threshold suggests that
at 1\,500~K these  sample about 85~\% of the total
contribution. This means that, 10to10
remains incomplete and that for higher temperatures an even  larger
number of lines will be required for accurate modeling of the opacity of
methane in hot media such as exoplanets and (cool) stars.





\begin{acknowledgments}
  This work was supported by the UK Science and Technology Resrearch
  Council (STFC), ERC Advanced Investigator Project 267219 and the
  Australian Research Council through Discovery grant DP110103167.
  This work made extensive use of the DiRAC@Darwin and DiRAC@COSMOS HPC
  clusters.  DiRAC is the UK HPC facility for particle physics,
  astrophysics and cosmology which is supported by STFC and BIS.  We
  also thank UCL for use of the Legion High Performance Computer for
  performing the electronic structure calculations.
\end{acknowledgments}



n




\end{article}

\end{document}